\documentclass[review=false]{elsarticle}

\usepackage{lineno,hyperref}
\usepackage{times}
\usepackage{latexsym}
\usepackage{graphicx,subfigure}
\usepackage{epstopdf}
\usepackage{amsmath}
\usepackage{breqn, algorithm, algpseudocode}
\usepackage{multirow}

\modulolinenumbers[5]

\journal{Journal of \LaTeX\ Templates}

\begin{document}

\begin{frontmatter}

\title{STTM: A Tool for Short Text Topic Modeling}

\author{Jipeng Qiang$^{1,*}$}
\ead{jpqiang@yzu.edu.cn}
\author{Yun Li$^{1,*}$}
\ead{liyun@yzu.edu.cn}
\author{Yunhao Yuan$^{1}$}
\ead{yhyuan@yzu.edu.cn}
\author{Wei Liu$^{1}$}
\ead{weiliu@yzu.edu.cn}
\author{Xindong Wu$^{2}$}
\ead{xwu@louisiana.edu}
\address{$^1$Department of Computer Science, Yangzhou University, China}

\address{$^2$School of Computing and Informatics, University of Louisiana at Lafayette, USA}

\cortext[mycorrespondingauthor]{Corresponding author}





\begin{abstract}

Along with the emergence and popularity of social communications on the Internet, topic discovery from short texts becomes fundamental to many applications that require semantic understanding of textual content. As a rising research field, short text topic modeling presents a new and complementary algorithmic methodology to supplement regular text topic modeling, especially targets to limited word co-occurrence information in short texts. This paper presents the first comprehensive open-source package, called STTM, for use in Java that integrates the state-of-the-art models of short text topic modeling algorithms, benchmark datasets, and abundant functions for model inference and evaluation. The package is designed to facilitate the expansion of new methods in this research field and make evaluations between the new approaches and existing ones accessible. STTM is open-sourced at https://github.com/qiang2100/STTM.he

\end{abstract}


\end{frontmatter}


\section{Introduction}

Along with the emergence and popularity of social communications (e.g. Twitter and Facebook), short text has become an important information source. Inferring topics from the overwhelming amount of short texts becomes a critical but challenging task for many content analysis tasks \cite{cheng2014btm,wang2015exploring}. Existing traditional methods for long texts (news reports or papers) such as probabilistic latent semantic analysis (PLSA) \cite{hofmann1999probabilistic} and latent Dirichlet allocation (LDA) \cite{blei2003latent} cannot solve this problem very well since only very limited word co-occurrence information is available in short texts. Therefore, short text topic modeling has already attracted much attention from the machine learning research community in recent years, which aims at overcoming the problem of sparseness in short texts. 

As a rising research field, short text topic modeling presents a new and complementary algorithmic methodology to supplement topic modeling algorithms for long text, especially targets to limited word co-occurrence information in short texts. To the best of our understanding, there is no comprehensive open-source library available for short text topic modeling algorithms. To facilitate the research on mining latent topics from short texts, we present a novel software package called STTM (Short Text Topic Modeling). The main contribution of STTM lies on three aspects. (1) STTM is the first comprehensive open-source library, which not only includes the state-of-the-art algorithms with a uniform easy-to-use programming interface but also includes a great number of designed modules for the evaluation and application of short text topic modeling algorithms. (2) STTM includes traditional topic modeling algorithms for long texts, which can be conveniently compared with short text topic modeling. (3) STTM is written in Java, easy to use and completely open source. Therefore, new approaches are easily integrated and evaluation through the STTM framework. The package STTM is available at https://github.com/qiang2100/STTM.

\section{Design Principles and System Architecture}

The framework of STTM follows three basic principles. (1) Preferring integration of existing algorithms rather than implementing them. If the original implementations of algorithms are open, we always attempt to integrate the original codes rather than implement them. The work that we have done is to consolidate the input/output file formats and package these different approaches into some newly designed java classes with a uniform easy-to-use member functions. (2) Including traditional topic modeling algorithms for long texts. The classical topic modeling algorithms (LDA \cite{blei2003latent} and its variation LF-LDA \cite{nguyen2015improving}) are integrated, which is easy for users to comparison of long text topic modeling algorithms and short text topic modeling algorithms. (3) Extendibility. Because short text topic modeling is an emerging research field, many topics such as hierarchical variational models for short texts have not been studied yet. For incorporating future work easily, we try to make the class structures as extendable as possible when designing the core modules of STTM.

\begin{figure}
	\centering
	\includegraphics[scale=0.6]{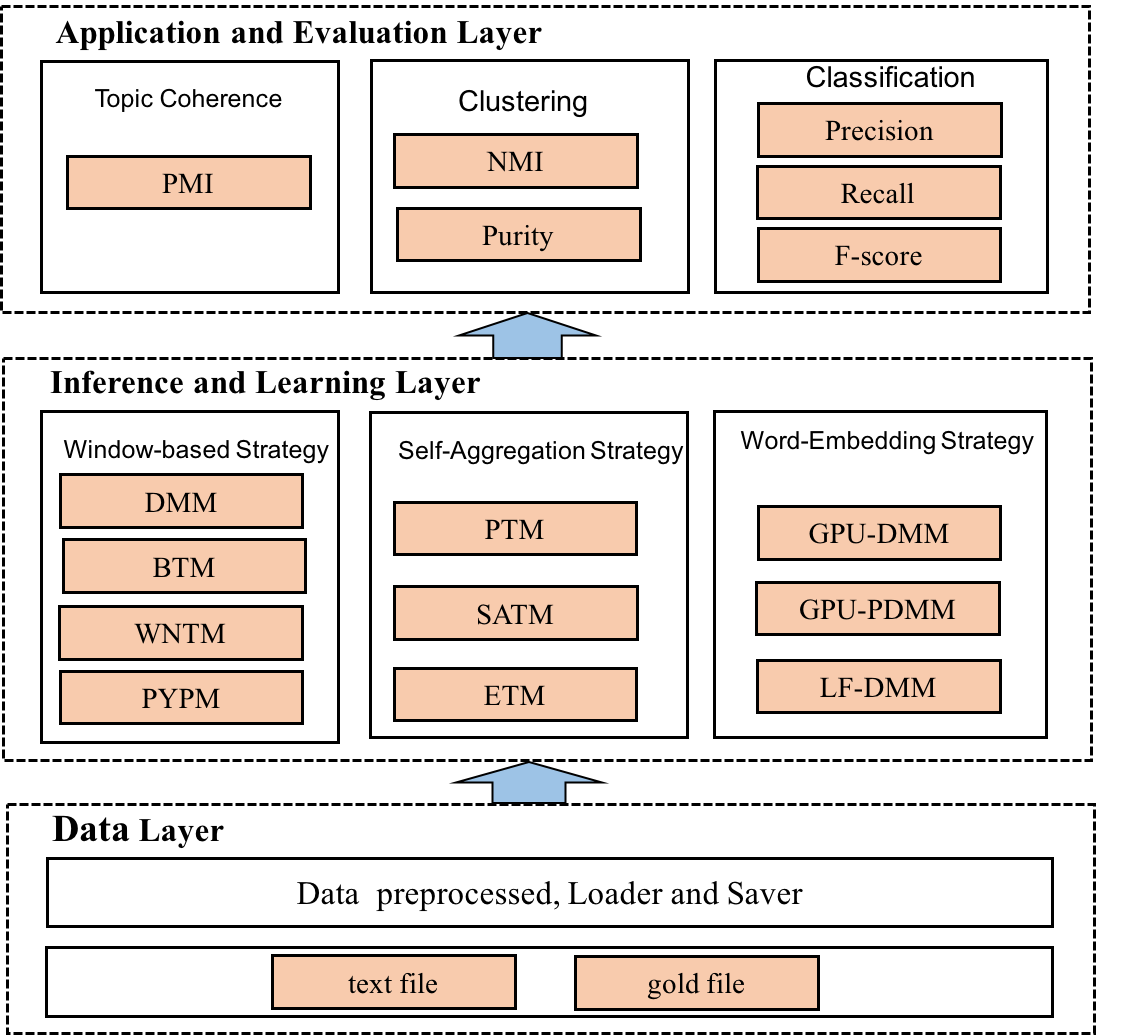}
	\caption{The architecture of STTM}
\end{figure}

Figure 1 shows the hierarchical architecture of STTM. STTM supports the entire knowledge discovery procedure including analysis, inference, evaluation, application for classification and clustering. In the data layer, STTM is able to read a text file, in which each line represents one document. Here, a document is a sequence of words/tokens separated by whitespace characters. If we need to evaluate the algorithm, we also need to read a gold file, in which each line is the class label of one document. In the inference and learning layer, STTM includes a large number of short text topic modeling algorithms. For each model, we not only provide how to train a model on existing corpus but also give how to infer topics on a new/unseen corpus using a pre-trained topic model. For alleviating the problem of sparseness, there are the following three main heuristic strategies in short text topic modeling: window strategy, self-aggregation strategy, and word-embedding strategy. In the application and evaluation layer, STTM presents three aspects about how to evaluate the performance of the algorithms (i.e., topic coherence, clustering, and classification). For topic coherence, we use the point-wise mutual information (PMI) to measure the coherence of topics \cite{zuo2016topic}. Short text topic modeling algorithms are widely used for clustering and classification. In the clustering module, STTM provides two measures (NMI and Purity) \cite{yan2015probabilistic}. Based on the latent semantic representations learned by short text topic modeling, three measures (macro averaged precision, recall and f-score) are used in classifications \cite{cheng2014btm}.

\section{Algorithms}

Due to only very limited word co-occurrence information in short texts, how to extract topics from short texts remains a challenging research problem \cite{wang2015exploring}. Three major heuristic strategies have been adopted to deal with how to discover the latent topics from short texts. One follows window-based strategy that two words or all words in one window are sampled from only one latent topic which is totally unsuited to long texts, but it can be suitable for short texts compared to the complex assumption that each text is modeled over a set of topics \cite{yan2015probabilistic,zhao2011comparing}. Therefore, many models (DMM\cite{yin2014dirichlet}, BTM\cite{cheng2014btm}, WNTM\cite{zuo2016word} and PYPM \cite{qiang2018short}) for short texts were proposed based on this window-based strategy. The second strategy aggregates short texts into long pseudo-texts before topic inference that can help improve word co-occurrence information. In this framework, STTM based self-aggregation strategy includes three algorithms, PTM \cite{zuo2016topic}, SATM \cite{quan2015short} and ETM \cite{qiang2017topic}. The last scheme directly leverages recent results by word embeddings that obtain vector representations for words trained on very large corpora to improve the word-topic mapping learned on a smaller corpus. Using word-embedding strategy, STTM integrates the following algorithms, GPU-DMM\cite{li2016topic},GPU-PDMM \cite{li2017enhancing} and LF-DMM \cite{nguyen2015improving}, which are the variations of DMM by incorporating the knowledge of word embeddings.

\section{Usage Example}

STTM can be easily executed in Linux and Windows systems. The STTM library comes with detailed documentation \footnote{https://github.com/qiang2100/STTM}. Figure 2 gives a sample example about training short text topic modeling from STTM. In this sample example, all models provide a uniform interface function, which can be easily set the parameters for each model. 

\begin{figure*}
	\centering
	\includegraphics[scale=0.6]{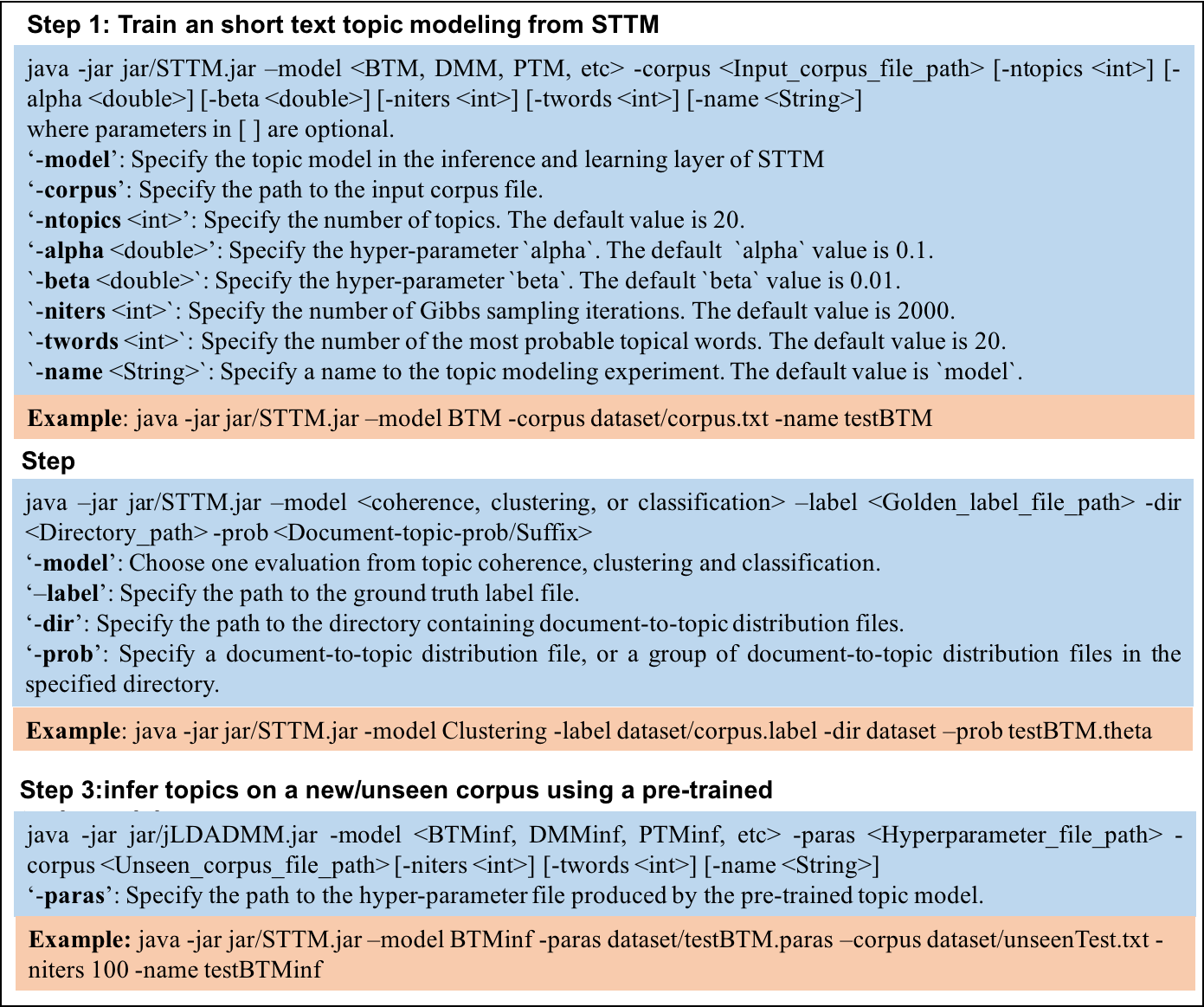}
	\caption{A sample example for a basic usage}
\end{figure*}

In the first step, we need to select one algorithm from all short text topic modeling algorithms to learn the latent topics of short texts. Because each algorithm has their self-parameters, STTM provides the options which can be specified by users. Below each step, we give one example about training a model using BTM algorithm. After this step using BTM, the output files including 'testBTM.theta', 'testBTM.phi', `testBTM.topWords`, `testBTM.topicAssignments` and `testDBTM.paras` are generated in the `dataset` folder. In the second step, we can evaluate the performance of the used model in step 1 by choosing one evaluation from topic coherence, clustering and classification. If we choose clustering in this step, the results of NMI and Purity are showed in one file "testBTM.theta.PurityNMI". If you need to infer topics for a new/unseen corpus, you can execute the third step.

\section{Conclusion and Future Work}
STTM is an easy-to-use open-source library for short text topic modeling to facilitate research efforts in machine learning and data mining. The recent version of STTM is consisted of many short text topic modeling algorithms, unseen corpus inferring and beneficial modules supporting different evaluations. Through the STTM framework, we hope that programmers or researchers can easily use short text topic modeling algorithms to discover latent topics fro short texts, and new methods are easily integrated into STTM framework and evaluation.

\section*{Acknowledgement}

This research is partially supported by the National Natural Science Foundation of China under grants (61703362, 61702441, 61402203), the Natural Science Foundation of Jiangsu Province of China under grants (BK20170513, BK20161338), the Natural Science Foundation of the Higher Education Institutions of Jiangsu Province of China under grant 17KJB520045, and the Science and Technology Planning Project of Yangzhou of China under grant YZ2016238.

\section*{References}

\bibliography{STTM}

\end{document}